\documentclass[twocolumn]{aastex62}

\usepackage{amsmath}
\usepackage{amssymb}
\usepackage{color}
\usepackage{mathrsfs}
\usepackage{enumerate}
\newcounter{address}
\def\vector#1{\mbox{\boldmath $#1$}}
\newcommand{\pc}{\ensuremath{\,\mathrm{pc}}}
\newcommand{\kpc}{\ensuremath{\,\mathrm{kpc}}}
\newcommand{\Myr}{\ensuremath{\,\mathrm{Myr}}}

\newcommand{\kms}{\ensuremath{\,\mathrm{km\ s}^{-1}}}
\newcommand{\kmskpc}{\ensuremath{\,\mathrm{{km\ s}^{-1}\ {kpc}^{-1}}}}
\newcommand{\mas}{\ensuremath{\,\mathrm{mas}}}
\newcommand{\masyr}{\ensuremath{\,\mathrm{mas\ yr}^{-1}}}
\newcommand{\microas}{\ensuremath{\,\mu\mathrm{as}}}
\newcommand{\microasyr}{\ensuremath{\,\mu\mathrm{as\ yr}^{-1}}}
\newcommand{\vlos}{v_{\ensuremath{\mathrm{los}}}}
\newcommand{\muell}{\mu_{\ell*}}
\newcommand{\mub}{\mu_{b}}
\newcommand{\Teff}{T_{\ensuremath{\mathrm{eff}}}}
\newcommand{\logg}{\log g}
\newcommand{\flag}[1]{\texttt{\lowercase{#1}}}
\newcommand{\emcee}{\flag{emcee}}
\newcommand{\PyGaia}{\texttt{{PyGaia}}}

\newcommand{\Gaia}{\textit{Gaia}}
\newcommand{\eq}[1]{\begin{align}#1\end{align}}
\newsavebox\myVerb

\newcommand{\hvs}{\object[2MASS J09120652+0916216]{LAMOST-HVS1}}
\begin{document}

\title{Constraining Solar position and velocity with a Nearby Hypervelocity Star}

\correspondingauthor{Kohei~Hattori}
\email{khattori@umich.edu}

\author[0000-0001-6924-8862]{Kohei Hattori}
\affil{Department of Astronomy, University of Michigan, Ann Arbor, MI, 48104, USA}
\author[0000-0002-6257-2341]{Monica~Valluri}
\affil{Department of Astronomy, University of Michigan, Ann Arbor, MI, 48104, USA}
\author{Norberto~Castro}
\affil{Department of Astronomy, University of Michigan, Ann Arbor, MI, 48104, USA}
\affil{Leibniz-Institut f\"{u}r Astrophysik Potsdam (AIP), An der Sternwarte 16, 14482, Potsdam, Germany}

\begin{abstract}

Gravitational 3-body interaction among binary stars and the supermassive black hole (SMBH) at the center of the Milky Way 
occasionally ejects a hypervelocity star (HVS) with a velocity of $\sim1000 \kms$. 
Due to the ejection location, 
such a HVS initially has negligible azimuthal angular momentum $L_z \simeq 0 \kpc\kms$. 
Even if the halo is mildly triaxial, $L_z$ of a recently ejected nearby HVS remains negligible, 
since its flight time from the Galactic Center is too short to accumulate noticeable torque. 
However, if we make a wrong assumption about the Solar position and velocity, 
such a HVS would apparently have noticeable non-zero 
azimuthal  
angular momentum, due to the wrong reflex motion of the Sun. 
Conversely, with precise astrometric data for a nearby HVS, 
we can measure the Solar position and velocity 
by
assuming 
that the HVS has zero
azimuthal  
angular momentum. 
Based on this idea, 
here we propose a method to estimate 
the Galactocentric distance of the Sun $R_0$ and the Galactocentric Solar azimuthal velocity $V_\odot$ 
by using a HVS. 
We demonstrate with mock data for a  nearby HVS candidate that 
the \Gaia\ astrometric data, 
along with the currently available constraint on $V_\odot/R_0$ from the proper motion measurement of Sgr A*, 
can constrain $R_0$ and $V_\odot$ with uncertainties of $\sim 0.27 \kpc$ and $\sim 7.8 \kms$ (or fractional uncertainties of $3\%$), respectively. 
Our method will be a promising tool to constrain $(R_0, V_\odot)$, 
given that \Gaia\ is expected to discover many nearby HVSs in the near future. 
\end{abstract}
\section{Introduction} \label{section:introduction}

Precise measurement of the position $\vector{r}_\odot$ and velocity $\vector{v}_\odot$ of the Sun 
with respect to the Galactic Center (or in the Galactic rest frame) is  
key to understanding the dynamics of stars in the Milky Way. 
This is simply because what we observe 
is the heliocentric position $\vector{r}_{\rm helio}$ and velocity $\vector{v}_{\rm helio}$ of stars 
with respect to the Sun (or observer); 
while what we need to know is the 3D position 
$\vector{r} = \vector{r}_\odot + \vector{r}_{\rm helio}$
and 3D velocity 
$\vector{v} = \vector{v}_\odot + \vector{v}_{\rm helio}$
of stars in the Galactic rest frame in order to describe the motions of stars in the Milky Way. 
In particular, 
$R_0$ (the Galactocentric distance of the Sun) is the most uncertain quantity that characterizes $\vector{r}_\odot$, 
and 
$V_\odot$ (azimuthal velocity of the Sun in the Galactic rest frame) is the most uncertain component of $\vector{v}_\odot$, 
and many authors have proposed and used various methods to estimate these quantities
(recent examples include 
\citealt{Gillessen2017}, \citealt{Malhan2017}, and \citealt{Hayes2018}). 
Currently,
typical fractional uncertainty in $R_0$ in the literature is $\simeq 5\%$ 
(see table 3 of \citealt{BlandHawthorn2016}). 
Given that the astrometric satellite \Gaia\ \citep{GaiaCollaboration2016} will provide accurate measurements of  
$\vector{r}_{\rm helio}$ and $\vector{v}_{\rm helio}$ of stars, 
more accurate measurements of $R_0$ and $V_\odot$ 
which will enhance the value of \Gaia\ data 
for various other studies of Galactic structure and dynamics.

Currently, 
one of the most reliable constraints
on $(R_0, V_\odot)$ comes from the measurement 
of the proper motion of Sgr A*, the radio source associated with the supermassive black hole (SMBH) at the very center of the Milky Way. 
Since Sgr A* is thought to be at rest in the Galactic Center, 
its proper motion along the Galactic longitude reflects the angular motion of the Sun in the Galactic disc plane. 
This proper motion was measured with $0.4\%$ precision by \cite{Reid2004}, 
therefore the ratio $V_\odot / R_0$ is constrained with $0.4\%$ precision. 
Given that $V_\odot / R_0$ is tightly constrained, we only need to add one more independent constraint on $(R_0, V_\odot)$ 
to measure $R_0$ and $V_\odot$ individually; 
and here we propose a new dynamical method which makes use of hypervelocity stars (HVSs).

Orbits of stars within $0.5$ arcsec ($\simeq 0.02 \pc$) of the Galactic Center 
have unambiguously revealed the existence of a central SMBH 
with a mass of $M=4 \times 10^6 M_\odot$ \citep{Eisenhauer2003,Ghez2008,Gillessen2009,Gillessen2017}. 
\cite{Hills1988} theoretically predicted that 
when a tightly bound stellar binary passes close enough to this SMBH, 
the strong tidal force can disrupt the stellar binary. 
The result of this 3-body interaction 
is that 
one of the binary stars (the so-called S-star) is decelerated and gravitationally captured by the SMBH, 
while 
the other star attains kinetic energy and escapes from the central region of the Milky Way. 
The typical ejection velocity of this escaping star, or the 
HVS, 
is of the order of $10^3 \kms$, 
exceeding the escape velocity at the Galactic Center. 
There are several possible definitions for HVSs \citep{Brown2015ARAA}, 
but in this paper we adopt the following definition:
\begin{itemize}
\item[] We define HVSs as those stars which are ejected from the center of the Milky Way. 
\end{itemize}
We adopt this definition 
since our method which we apply to nearby stars 
is not sensitive to whether or not a star is bound to the Milky Way 
so long as it is ejected 
recently 
(less than a few orbital periods) 
from the Galactic Center. 
For example, 
a short-lived massive star with large space motion 
is a good candidate for such a star.

After the first discovery of HVS candidate by \cite{Brown2005}, 
about 20 HVS candidates have been reported so far \citep{Brown2014, Brown2015ApJ,Zheng2014,Huang2017}. 
We expect that \Gaia\ will discover more HVS candidates \citep{Marchetti2018}, 
and that some of the HVS candidates will be confirmed to be Galactic Center origin with precise \Gaia\ astrometry.

If a HVS is ejected from the Galactic Center by the Hills mechanism, 
the magnitude of its azimuthal angular momentum $|L_z|$ is much smaller than 
that of the Sun, $|L_{z,\odot}| =  |R_0 V_\odot| \simeq 2\times 10^3 \kpc\kms$. 
The tidal breakup of a stellar binary near the Galactic Center typically happens at a radius of $\sim 30 \;\mathrm{AU}$ 
and the ejection velocity of a HVS is typically $\sim 10^3 \kms$. 
Thus, the typical value of $|L_z|$ of a HVS just after the ejection is $|L_z| \sim 1.5 \times 10^{-4} \kpc\kms$. 
In the region where the Galactic potential is nearly axisymmetric, 
the angular momentum $L_z$ is nearly conserved along the orbit. 
Also, even if the potential is mildly triaxial, 
$L_z$ is nearly zero for a recently ejected nearby HVS, 
since it does not have long enough flight time to accumulate torque from the potential and acquire noticeable non-zero $L_z$. 
As a result, 
recently ejected
HVSs that are relatively close to the Sun are expected to have negligible $L_z$. 
Therefore, if the heliocentric position $\vector{r}_{\rm helio}$ and velocity $\vector{r}_{\rm helio}$ of a nearby HVS 
can be accurately measured with astrometric surveys like \Gaia,
we can constrain the position $\vector{r}_\odot$ and the velocity of the Sun $\vector{v}_\odot$ in the Galactic rest frame, 
such that the $z$-component of the angular momentum is zero, i.e., 
\eq{
\left[ (\vector{r}_\odot + \vector{r}_{\rm helio}) \times (\vector{v}_\odot + \vector{v}_{\rm helio}) \right]_z= 0, 
}
for this HVS. 
To the best of our knowledge, this idea of using HVSs to constrain $(\vector{r}_\odot, \vector{v}_\odot)$ or $(R_0, V_\odot)$ has never been explored.

In this paper, we provide a theoretical framework to constrain $(R_0, V_\odot)$ 
from astrometric observations of a single HVS. 
The outline of this paper is as follows.
In Section \ref{section:formulation}, we explain our Bayesian formulation. 
In Section  \ref{section:MockData}, we briefly describe a recently discovered HVS candidate, \hvs, and generate \Gaia-like mock data of this star with the assumption that this star originates from the Galactic Center. 
In Section \ref{section:results}, we apply our formulation to the mock data and assess the accuracy of our method. 
In Section \ref{section:discussion} we provide some discussion about this method, and Section \ref{section:conclusion} sums up.


\section{Formulation} \label{section:formulation}

Here we describe our Bayesian formulation 
to estimate $\theta_G \equiv (R_0, V_\odot)$ 
by using the astrometric data of a nearby HVS,
which is ejected from the Galactic Center.

\subsection{Coordinate system} \label{section:coordinate}

We define a right-hand Cartesian coordinate system centered at the Galactic Center 
in which $(x,y)$-plane corresponds to the Galactic plane and $z$ is directed toward the North Galactic Pole. 
We assume that the Sun is located at $(x,y,z)=(-R_0, 0, z_0)$ 
and has a velocity of $(v_x, v_y, v_z)=(U_\odot, V_\odot, W_\odot)$. 
We assume that $(z_0, U_\odot, W_\odot)$ are well determined a priori 
and $(R_0, V_\odot)$ are to be determined. 
The uncertainties in $z_0$ and $(U_\odot, W_\odot)$ are much smaller than those of $R_0$ and $V_\odot$, respectively, 
so our assumption is well justified.

In Section \ref{section:MockData}, 
we generate our mock data by assuming 
$(R_0, z_0) = (8 \kpc, 0 \pc)$ and 
$(U_\odot, V_\odot, W_\odot)=(11.1, 240, 7.25) \;{\rm km\;s^{-1}}$ 
(see \citealt{Schoenrich2010} for $U_\odot$ and $W_\odot$).

\subsection{Observables of a HVS}

Let us define the observed 6-dimensional phase-space vector of a (nearby) HVS as 
\eq{
\vector{q}=(DM, \ell, b, \vlos, \mu_{\ell*}, \mu_b). 
}
Here, 
$DM$ is the distance modulus, 
$(\ell, b)$ are respectively the Galactic longitude and latitude (which are assumed to be determined with infinite precision), 
$\vlos$ is the heliocentric line-of-sight velocity, and 
$(\mu_{\ell*}, \mu_b)$ are respectively the proper motions in the $\ell$- and $b$-directions.
Also, we define the true 6-dimensional phase-space vector of this HVS,
\eq{
\vector{q}'=(DM', \ell, b, \vlos', \muell', \mub'). 
}
such that $\vector{q}$ and $\vector{q}'$ are identical in the limit of no observational error. 

In addition, we define a 4-dimensional vector that expresses the observational error given by 
\eq{
\vector{\sigma}_q = (\sigma_{DM}, \sigma_{v}, \sigma_{\mu \ell*}, \sigma_{\mu b}). 
} 
Here, $(\sigma_{DM}, \sigma_{v}, \sigma_{\mu \ell*}, \sigma_{\mu b})$ 
are the Gaussian errors in $(DM, \vlos, \mu_{\ell*}, \mu_b)$, respectively. 
In this paper, we simply assume $\sigma_{\mu \ell*} = \sigma_{\mu b} = \sigma_\mu$, 
but our formulation can be easily generalized to 
include the correlation coefficient $\rho_{\ell b}$ ($-1 \leq \rho_{\ell b} \leq 1$) 
in the error distribution of $(\muell, \mub)$.

\subsection{$L_z$ of a HVS}

The azimuthal angular momentum $L_z=(x v_y - y v_x)$ of a HVS is essentially zero 
-- since it originates from the Galactic Center.
Therefore, 
the probability distribution function of the true astrometric coordinates 
$(DM', \vlos', \muell', \mub')$ of a HVS
given the parameters $\theta_G = (R_0, V_\odot)$ and the observed direction $(\ell, b)$ 
can be simply assumed to be a Dirac Delta function 
\eq{ \label{eq:pdf_simple}
P(DM', \vlos', \muell', \mub' | \ell, b, \theta_G ) 
= E \delta ( L_z( \vector{q}', \theta_G ) ) 
}
with $E$ a constant and 
\eq{
L_z( \vector{q}', \theta_G ) = A \vlos' + B \mu'_{\ell*} + C \mu'_b + D .
}
Here, $A$, $B$, $C$, and $D$ are given by
\eq{
\begin{cases}
&A= - R_0 \cos b \sin \ell \\
&B= k d' (d' \cos b - R_0 \cos \ell) \\
&C= k d' R_0 \sin b \sin \ell \\
&D= d' \cos b (\cos \ell V_\odot - \sin \ell U_\odot) - R_0 V_\odot ,
\end{cases}
}
$k=4.74047 \kms \kpc^{-1} \mas^{-1}$ is a constant, 
and $d'$ is the distance that corresponds to $DM'$. 

Of course, with a detailed model of ejection mechanism of HVSs (e.g., \citealt{Yu2007, Kenyon2008, Rossi2017, Marchetti2018}), 
we can think of more sophisticated probability distribution 
than equation (\ref{eq:pdf_simple}). 
However, such sophistication may not be helpful for our problem,
since we shall use only a single HVS.

\subsection{Likelihood function}

The probability 
that a HVS is found at $\vector{q}$ in the observable space 
given the Galactic parameters $\theta_{\rm G}=( R_0, V_\odot )$ 
and the observational errors $\vector{\sigma}_q$ can be expressed as 
\eq{
&P(\vector{q} | \theta_{\rm G}, \vector{\sigma}_q ) 
= \int d \vector{q}' \; P( \vector{q} | \vector{q}' , \vector{\sigma}_q ) P( \vector{q}' | \theta_{\rm G})\\
&= \int d \vector{q}' \; P( \vector{q} | \vector{q}' , \vector{\sigma}_q ) 
P(DM', \vlos', \muell', \mub' | \ell, b, \theta_G ) P(\ell, b | \theta_G).
}
If the observational errors in $(DM, \vlos, \muell, \mub)$ are Gaussian, 
then we have
\eq{
&P(\vector{q} | \theta_{\rm G}, \vector{\sigma}_q ) \\
&= \frac{P(\ell, b | \theta_G) E}{(2 \pi)^{2} \sigma_{DM} \sigma_v \sigma_\mu \sigma_\mu } 
\iiiint_{-\infty}^{\infty} d DM' d \vlos' d \mu'_\ell d \mu'_b \; \nonumber \\
&\times \delta ( L_z( \vector{q}', \theta_G ) )  \exp \left[ 
 - \frac{(DM - DM')^2}{2 \sigma^2_{DM}} \right] \nonumber \\
&\times \exp \left[ - \frac{(\vlos - \vlos')^2}{2 \sigma^2_v} 
- \frac{(\mu_\ell - \mu'_\ell)^2}{2 \sigma^2_\mu} 
- \frac{(\mu_b - \mu'_b)^2}{2 \sigma^2_\mu} \right] ,
}
which reduces to 
\eq{
&P(\vector{q} | \theta_{\rm G}, \vector{\sigma}_q ) \nonumber \\
&= \frac{P(\ell, b | \theta_G) E}{2 \pi \sigma_{DM} } 
\int_{-\infty}^{\infty} d DM' \; \exp \left[  - \frac{(DM - DM')^2}{2 \sigma^2_{DM}} \right] \nonumber \\
&\times \frac{1}{\sqrt{A^2 \sigma_v^2 + B^2 \sigma_\mu^2 + C^2 \sigma_\mu^2 }} \nonumber \\
&\times \exp \left[ - \frac{( A \vlos + B \mu_\ell + C \mu_b + D )^2}
{2(A^2 \sigma_v^2 + B^2 \sigma_\mu^2 + C^2 \sigma_\mu^2) } 
\right] . \label{eq:likelihood}  
}
In this paper, we assume a Gaussian error distribution for $DM$ for simplicity, 
but the likelihood function in equation (\ref{eq:likelihood}) 
can be easily generalized to a non-Gaussian error distribution for $DM$ (see Appendix \ref{appendix:nonGaussianDM}).

\subsection{Prior distribution}\label{section:prior}

The proper motion of Sgr A* 
is observed to be $\mu_{\ell*, \mathrm{SgrA}} = 6.379 \pm 0.024 \masyr$ \citep{Reid2004}. 
If Sgr A* is at rest at the Galactic Center, 
this proper motion solely arises from the Solar reflex motion, 
and we have 
$V_\odot/R_0 = k \mu_{\ell*, \mathrm{SgrA}} = 30.24 \pm 0.114 \kmskpc$. 
Thus, 
we adopt a prior distribution  for $\theta_{\rm G} = ( R_0, V_\odot )$ of the form 
\eq{ 
P( R_0, V_\odot ) dR_0 dV_\odot
\propto dR_0 dV_\odot 
\exp \left[ - \frac{(V_\odot/R_0 - \Omega)^2}{2 \sigma^2_{\Omega}} \right] . \label{eq:prior}
}
In our mock analyses in Section \ref{section:results}, 
we adopt $\Omega = V_\odot / R_0 = 30 \; \kmskpc$, 
and $\sigma_\Omega=0.116 \; \kms\kpc^{-1}$, 
by assuming that the Brownian motion of Sgr A* 
with respect to the Galactic Center is about $0.2 \kms$ \citep{Merritt2007}.\footnote{
We note $\sigma_\Omega / \Omega = \sqrt{ (0.024/6.379)^2 + (0.2 \kms/V_\odot)^2}$.
}

\subsection{Bayesian formulation}

From Bayes theorem, 
the posterior probability distribution of $\theta_{\rm G}$ 
given the observed data $\vector{q}$ 
and information on observational error $\vector{\sigma}_{q}$ for HVSs 
is expressed as 
\begin{align}
P(\theta_{\rm G} | \vector{q}, \vector{\sigma}_{q}) = \frac{P(\vector{q}|\theta_{\rm G}, \vector{\sigma}_{q}) P(\theta_{\rm G})}{P(\vector{q} | \vector{\sigma}_{q})}. 
\end{align}
Here, 
$P(\theta_{\rm G})$ is the prior given in equation (\ref{eq:prior});
$P(\vector{q} | \vector{\sigma}_{q})$ is the evidence (which can be regarded as a constant); 
and the likelihood $P(\vector{q}|\theta_{\rm G}, \vector{\sigma}_{q})$ is given in equation (\ref{eq:likelihood}).


\begin{table}
\caption{Basic properties of LAMOST-HVS1}
\label{table:LAMOST-HVS1}
\centering
\begin{tabular}{cc}
	\hline\hline
	Observed quantities &   \\
	\hline
	$(\ell, b)$ & $(221.0996^\circ, 35.4072^\circ)$ \\
	$(\alpha, \delta)$ & $(138.033^\circ, 9.280^\circ)$ \\
	$B$ & $12.936 \pm 0.036$ \\
	$V$ & $13.055 \pm 0.009$ \\
	$J$ & $13.357 \pm 0.028$ \\
	$H$ & $13.43 \pm 0.04$ \\
	$K_s$ & $13.53 \pm 0.04$ \\
	(\Gaia\ DR1) $G$ & $14.677731837262346$ \\
	(Line-of-sight velocity)$^\dagger$ $v_{\rm los} \pm \sigma_v$ & $611.65 \pm 4.63 \kms$ \\
	\hline\hline
	Estimated quantities &   \\
	\hline
	$\log (g)$ & $3.5 \pm 0.1$ \\ 
	$T_\mathrm{eff}$ &  $20000 \pm 1000 \;\mathrm{K}$ \\ 
	age $\tau$ & $29.91^{+5.28}_{-6.62} \Myr$ \\
	(Current mass) $M_{*}$ & $9.17^{+1.47}_{-0.73} M_\odot$ \\
	(Current radius) $R_{*}$ & $9.02^{+1.59}_{-1.40} R_\odot$ \\	
	(Distance modulus) $DM \pm \sigma_{DM}$ & $16.276 \pm 0.569$ \\
	(Heliocentric distance) $d$ & $18.00^{+5.59}_{-4.03} \kpc$ \\
	\hline\hline
	Expected \Gaia\ error$^\ddagger$  &   \\
	\hline
	DR2 parallax error & $\sigma_{\varpi \mathrm{DR2}} = 30 \;\microas$ \\
	DR2 proper motion error & $\sigma_{\mu \mathrm{DR2}} = 43 \;\microasyr$ \\
	Final DR parallax error & $\sigma_{\varpi \mathrm{Final}} = 22.462 \;\microas$ \\
	Final DR proper motion error & $\sigma_{\mu \mathrm{Final}} = 11.818 \;\microasyr$ \\
	\hline
\end{tabular}
\begin{flushleft}
$^\dagger${
\cite{Huang2017}
}
$^\ddagger${
These quantities are estimated based on the $G$-band magnitude in \Gaia \ DR1.
}\\
\end{flushleft}
\end{table}

\section{Mock \Gaia\ data of LAMOST-HVS1} \label{section:MockData}

Our method of estimating $( R_0, V_\odot )$ from a single HVS can be applied to any HVS. 
Its performance is determined by the observational error on the position and velocity of the HVS. 
Thus, 
our method is most suited for nearby, bright HVSs 
for which accurate astrometric data can be obtained. 
With this regard, 
the recently discovered nearby HVS candidate, \hvs\ \citep{Zheng2014}, 
is a good test case to examine the accuracy of our method 
(but see also Section \ref{section:note}). 

In this section, we briefly review the current knowledge about \hvs. 
Then we generate mock \Gaia\ data for this star 
with the assumption that it was recently ejected from the Galactic Center. 
The mock data shall be used in Section \ref{section:results} to demonstrate the usefulness of our method. 
A simple test to check whether or not this star originates from the Galactic Center is described in Section \ref{section:PMdiagnosis}.

\subsection{Observed properties of LAMOST-HVS1} \label{section:observation}

The nearby HVS candidate, \hvs, was discovered by \cite{Zheng2014} through the LAMOST survey \citep{Cui2012, Zhao2012}; 
and this star is also discussed in \cite{Huang2017}. 
Table \ref{table:LAMOST-HVS1} shows some of its 
observed properties (e.g. position and magnitudes) as well as some properties inferred from the spectra.
Based on the $G$-band magnitude of this star in \Gaia\ Data Release 1 
(DR1;
\citealt{Lindegren2016}),
we expect that the parallax error and proper motion error in \Gaia\ DR2 will be  
about $\sigma_{\varpi \mathrm{DR2}} = 30 \;\mu\mathrm{mas}$ and $\sigma_{\mu \mathrm{DR2}} = 43 \;\microasyr$, respectively. 
Even if the stellar distance is as close as $d \simeq 13 \kpc$ \citep{Zheng2014}, 
the expected fractional error on parallax in \Gaia\ DR2 is about $\sigma_{\varpi \mathrm{DR2}} / (1/ 13 \kpc) = 0.39$, 
which is too large to determine the distance based on the parallax only \citep{Bailer-Jones2015}. 
Therefore, we need to measure stellar distance based on photometric and spectroscopic data. 

We re-analyzed the available LAMOST spectra\footnote{http://dr3.lamost.org/search}  using  a grid of synthetic atmosphere models built using the atmosphere/line formation code {\sc fastwind} \citep{1997A&A...323..488S,2005A&A...435..669P,2012A&A...543A..95R}. The observed spectra are compared with the synthetic {\sc fastwind} grid, retrieving the set of stellar parameters such as $(\Teff, \logg)$ that best reproduce the main chemical transitions in the data (details can be found in \citealt{2012A&A...542A..79C}). Current stellar radius, age and mass are estimated using the $(\Teff, \logg)$ and the rotating evolutionary tracks published by \cite{Ekstrom2012}. We estimated a distance modulus of $DM=16.276$ with an uncertainty of $\sigma_{DM}=0.569$ based on the stellar properties and photometry reported in Table \ref{table:LAMOST-HVS1}, adopting the standard \cite{1989ApJ...345..245C} extinction law. 
The corresponding heliocentric distance to this star is $d = 18.00^{+5.59}_{-4.03} \kpc$ 
(fractional distance error of $\simeq 27\%$). 
The heliocentric line-of-sight velocity $v_{\rm los}$ is adopted from \cite{Huang2017}.

In the above-mentioned analysis, we assume that this star is an ordinary sub-giant branch star. 
However, there is a possibility that this star is actually a $9 M_\odot$ blue straggler. 
Even if this star is a blue straggler formed out of a mass transfer or merger of binary stars \citep{Bailyn1995}, 
we expect that our age (after the formation of a blue straggler) and distance estimates are not seriously affected, 
as the evolution of blue straggler is quite similar to an ordinary star with the same mass \citep{Sills2009}.


\subsection{Construction of mock \Gaia\ data for LAMOST-HVS1} \label{section:data}

The currently available proper motion data for \hvs\ are associated with large error, 
and they will be superseded by \Gaia\ DR2 proper motion. 
Therefore, we generate mock \Gaia\ proper motion data based on 
the currently available quantities, $(DM, \sigma_{DM}, \ell, b, \vlos, \sigma_{v})$. 
The procedure to generate mock data is described in the following. 
We note that we generate two sets of mock data:
(a) 101 `controlled mock data' of a single mock HVS; and 
(b) 1000 `random mock data' of a single mock HVS.

\subsubsection{Controlled mock data} \label{section:control}

We generate 101 controlled mock data, 
each of which is identified by an integer $i$ ($i=1,2, \cdots, 101$). 
These controlled mock data are generated in the following manner: 

(Step 1) 
We assume that $R_0=8\kpc$, $z_0=0 \pc$ and $(U_\odot, V_\odot, W_\odot)=(11.1, 240, 7.25) \kms$ in our mock data, 
and that we know the exact values of $(z_0, U_\odot, W_\odot)$. 

(Step 2) 
Based on the observed values of $(DM, \vlos)$ and their associated errors of $(\sigma_{DM}, \sigma_v)$, 
we assign the `true' values of $(DM', \vlos')$. 
To be specific, we assign 
\eq{\label{eq:DM_i}
DM' = DM + (i - 51)/25 \times \sigma_{DM},
}
such that the true value of distance modulus increases as a function of $i$ (the mock data ID). 
By design, $i=26$ ($i=76$) corresponds to the case where the true distance modulus 
$DM'$
is smaller (larger) than the observed value $DM$ by $\sigma_{DM}$. 
Also, $i=51$ corresponds to the case where the true value $DM'$ happens to be the observed value $DM$.
In this manner, we can easily tell how the systematic error on distance affects the final result. 
For the true line-of-sight velocity, 
we randomly draw a velocity from a Gaussian distribution with mean $\vlos$ and dispersion $\sigma_v$:
\eq{
\vlos' \sim \mathcal{N}(\vlos, \sigma_v). 
}
Here, $\mathcal{N}(m, s)$ is a Gaussian distribution with mean $m$ and dispersion $s$.

(Step 3) 
Based on the true 4D information of $(DM', \ell, b, \vlos')$, 
we find the orbit that connects the current location and the Galactic Center, 
such that the star was located at the Galactic Center a few tens of $\Myr$ ago. 
Here, we adopt MWPotential2014 model \citep{Bovy2015} as the Galactic potential model.
After finding the orbit, we derive the corresponding current `true' proper motion $(\muell', \mub')$.

(Step 4) 
Finally, we add a random Gaussian noise 
that follows the expected proper motion error distribution for \Gaia\ DR2  
to the true proper motion $(\muell', \mub')$ 
and obtain the mock \Gaia\ proper motion data of $(\muell, \mub)$. 

\subsubsection{Random mock data} \label{section:random}

The 1000 random mock data are constructed in the same manner as the controlled mock data (Section \ref{section:control}), 
except that we draw the true distance modulus from a Gaussian distribution: 
\eq{
DM' \sim \mathcal{N}( DM,  \sigma_{DM} ),
}
instead of using equation (\ref{eq:DM_i}) in (step 2).


\section{Analyses of mock data} \label{section:results}

In this section we apply our Bayesian formulation (Section \ref{section:formulation}) to our 
\Gaia-like mock data described in Section \ref{section:data} 
to demonstrate how well we can recover the correct values of $(R_0^\mathrm{correct}, V_\odot^\mathrm{correct})=(8\;{\rm kpc}, 240\;{\rm km\;s^{-1}})$ 
with the astrometric data of a nearby HVS. 
In these analyses, we use MCMC package of \emcee\ \citep{Foreman-Mackey2013}. 

\subsection{Results for representative controlled mock data}

In order to evaluate how well our method performs, 
we first focus on three representative controlled mock data with $i =$ 26, 51, and 76. 
These mock data correspond to the cases where 
the true value of distance modulus $DM'$ is 
$DM-\sigma_{DM}$ ($i$: 26), 
$DM$ ($i$: 51), and 
$DM+\sigma_{DM}$ ($i$: 76).

\begin{figure}
\begin{center}
	\includegraphics[angle=0,width=0.8\columnwidth]{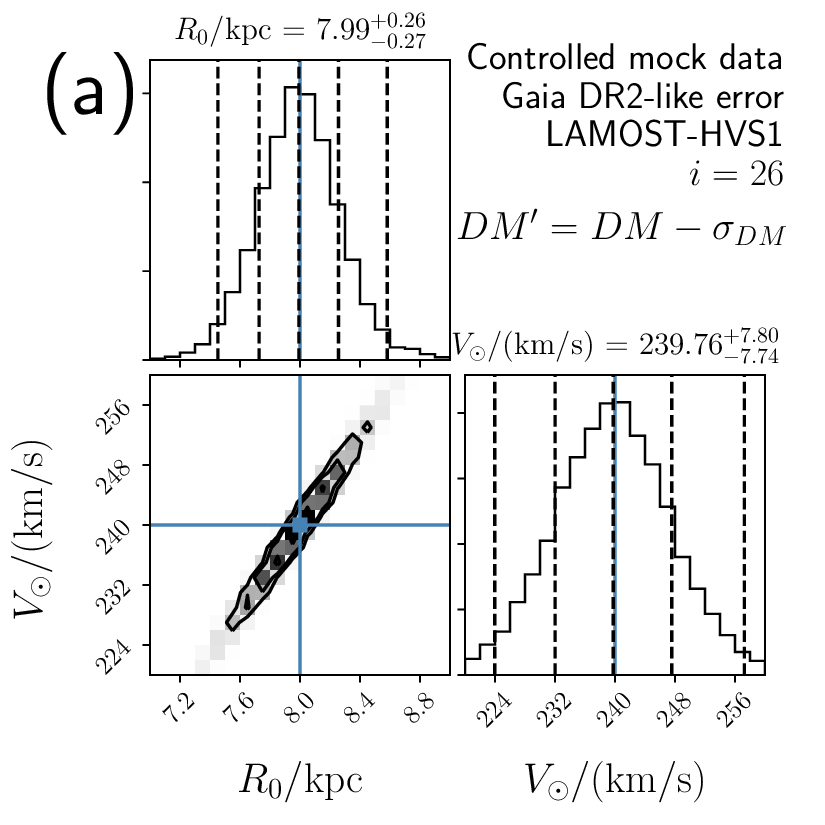}\\
	\includegraphics[angle=0,width=0.8\columnwidth]{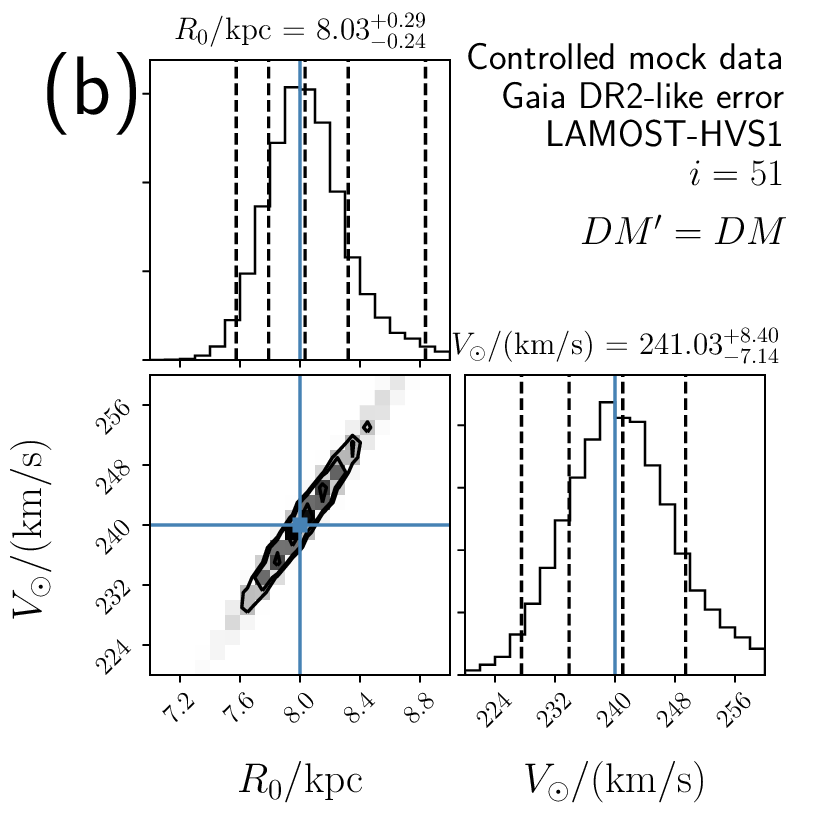}\\
	\includegraphics[angle=0,width=0.8\columnwidth]{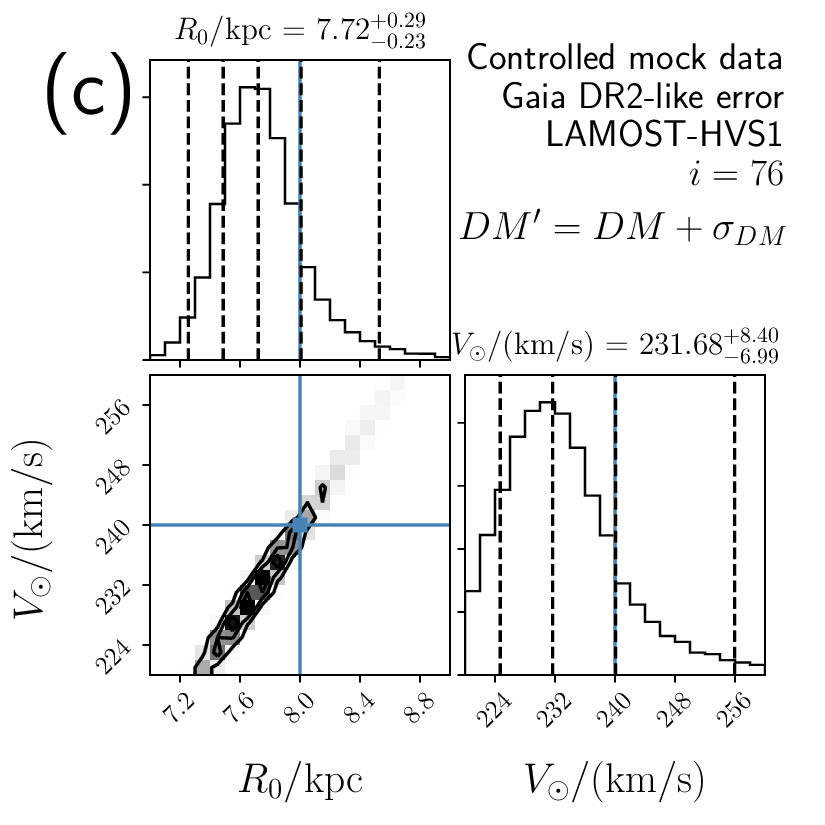}
\end{center}
\caption{
Posterior distributions for 
three representative controlled mock data of LAMOST-HVS1 with \Gaia\ DR2-like proper motion error. 
(a) 
Results for mock datum with $i=26$, where true distance modulus $DM'$ is expressed as $DM' = DM - \sigma_{DM}$. 
(b) 
The same figure but with $i=51$ and $DM'=DM$. 
(c) 
The same figure but with $i=76$ and $DM' = DM + \sigma_{DM}$. 
}
\label{fig:i26_51_76}
\end{figure}

Figure \ref{fig:i26_51_76} shows the two-dimensional posterior distribution of $(R_0, V_\odot)$ for these three mock data. 
In each panel, the blue horizontal or vertical lines correspond to the correct values of $(R_0^\mathrm{correct}, V_\odot^\mathrm{correct})$. 
In each histogram, the vertical dashed lines indicate either 2.5, 16, 50, 84, or 97.5 percentiles of the posterior distribution. 
As we can see, there is a tight correlation between $R_0$ and $V_\odot$, 
mainly due to the strong prior on $V_\odot/R_0$. 
We note that the correct values of $R_0$ and $V_\odot$ are within the central 68 percentiles (between 16 and 84 percentiles) 
for each of these mock data. 
The error bars for $R_0$ and $V_\odot$ are about $0.27 \kpc$ and $8 \kms$, respectively, 
so the fractional uncertainties of these parameters are about $3\%$. 
With some additional experiments, 
we find that these uncertainties can be improved 
by using a better distance (smaller $\sigma_{DM}$) or a better proper motion (smaller $\sigma_\mu$). 
Some details are described in Appendix \ref{appendix:distance}.

\subsection{Results for all the controlled mock data}

Given that \Gaia\ proper motion error will be as small as $\sigma_{\mu \mathrm{DR2}} = 43 \microasyr$ even with DR2, 
the largest source of uncertainty in our method comes from the stellar distance error. 
Therefore, it is fruitful to see how 
the difference between the true value of distance modulus $DM'$ and the observed value $DM$ affects our estimation. 
In our controlled mock data, $DM'$ is designed to be linearly dependent on the integer $i$ (mock data ID). 
This makes it easier for us to understand how $(DM' - DM)$ affects the posterior distribution.

\begin{figure*}
\begin{center}
	\includegraphics[angle=0,width=1.8\columnwidth]{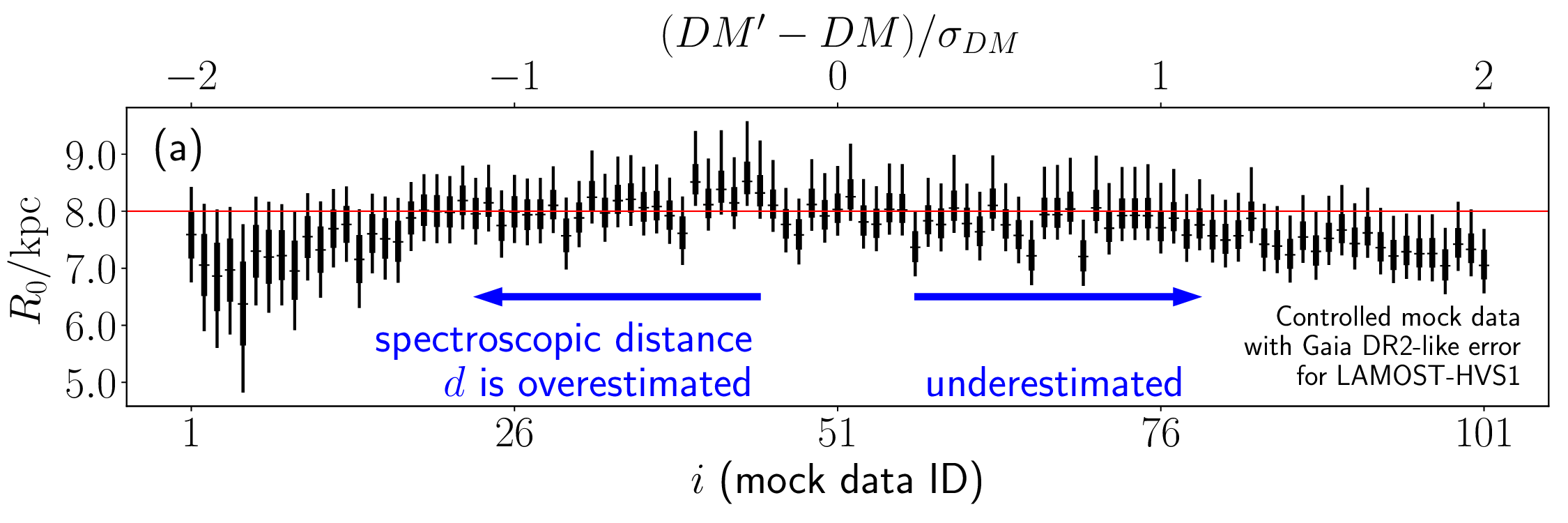}\\
	\includegraphics[angle=0,width=1.8\columnwidth]{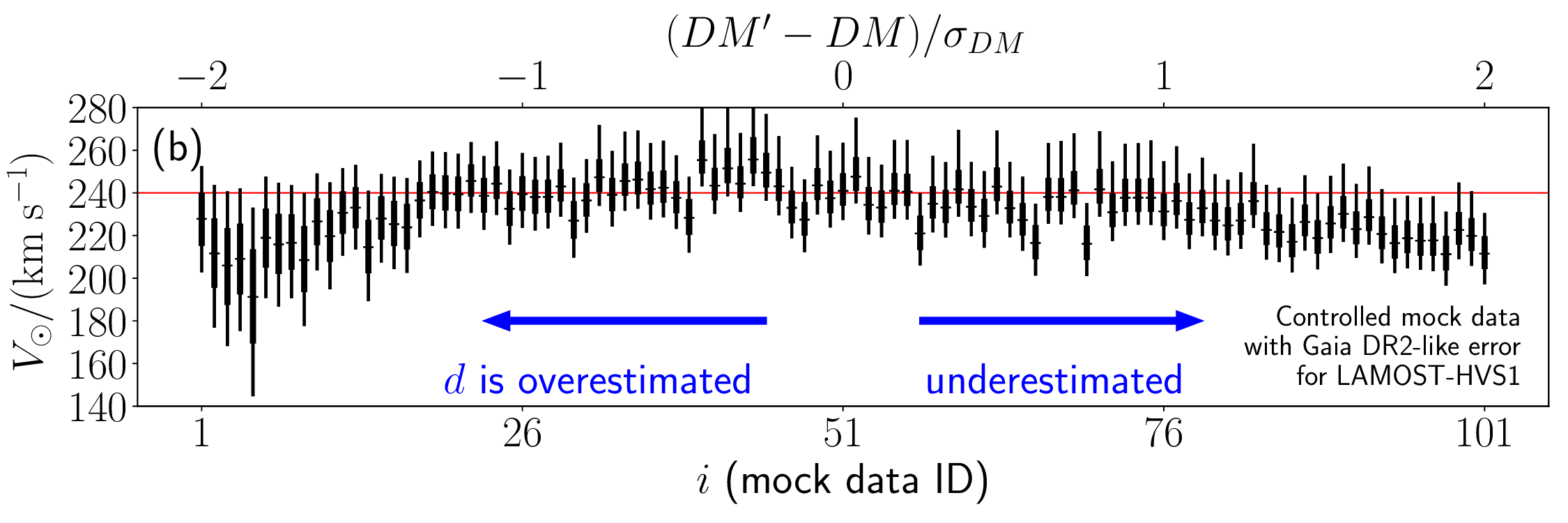}
\end{center}
\caption{
Posterior distributions for 
101 controlled mock data of LAMOST-HVS1 with \Gaia\ DR2-like proper motion error. 
Note that the true distance modulus $DM'$ is linearly dependent on $i$ as in equation (\ref{eq:DM_i}). 
Namely, the spectroscopic stellar distance $d$ is progressively more overestimated as $i$ decreases from 51 to smaller value;
while $d$ is progressively more underestimated as $i$ increases from 51 to larger. 
(a)
The posterior distributions of $R_0$. 
For each value of $i$, 
the thin vertical bar covers the 2.5 and 97.5 percentiles of the posterior distribution (central 95 percentiles), 
the thick vertical bar covers the 16 and 84 percentiles of the posterior distribution (central 68 percentiles), 
and the horizontal bar shows the 50 percentile (median). 
The horizontal red line indicates the correct value $R_0^\mathrm{correct}=8 \kpc$. 
(b)
The same as in (a), but for the posterior distributions for $V_\odot$. 
The horizontal red line indicates the correct value $V_\odot^\mathrm{correct}=240 \kms$. 
}
\label{fig:i_all}
\end{figure*}

Figure \ref{fig:i_all}(a) shows the posterior distribution of $R_0$ for our 101 controlled mock data as a function of $i$ and $(DM' - DM)$. 
We can see from this figure that the median values of the posterior distributions are located close to the correct value of $R_0^\mathrm{correct}$ 
when 
$-\sigma_{DM} \leq DM' - DM \leq \sigma_{DM}$ (when $26 \leq i \leq 76$). 
On the other hand, 
the posterior distributions are shifted toward lower $R_0$ (resulting in an underestimation of $R_0$) 
when 
$| DM' - DM | > 1.2 \sigma_{DM}$ (when $1\leq i \leq 20$ or $80 \leq i \leq 101$). 
We note that the posterior distributions are highly broadened for $-2 \sigma_{DM} \leq DM' - DM \leq -1.64 \sigma_{DM}$ ($i=1$-$10$), 
as can be seen from their elongated error bars in Figure \ref{fig:i_all}(a).
Fortunately, this region corresponds to $7.16 \leq d'/(\kpc) \leq 8.46$, 
where the fractional distance error based on the \Gaia\ DR2 parallax is smaller than 25\%. 
In such a case, we can use the parallax-based distance, which is more reliable than spectroscopic distance.

Figure \ref{fig:i_all}(b) shows the posterior distribution of $V_\odot$ as a function of $i$. 
Since we have applied a strong prior on $V_\odot/R_0$, 
the way the posterior distribution of $V_\odot$ depends on $i$ or $(DM'-DM)$ is very similar to that of $R_0$.

\subsection{Results for random mock data} \label{section:result_random}

In order to study the statistical robustness of our method, we analyzed the 1000 random mock data.

\begin{figure*}
\begin{center}
	\includegraphics[angle=0,width=0.7\columnwidth]{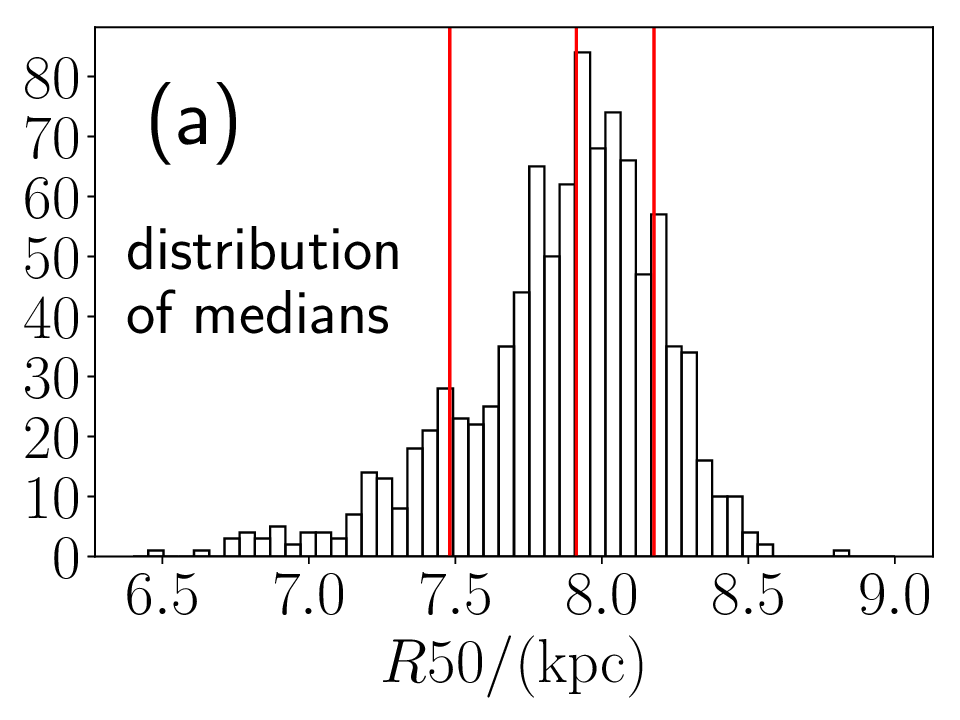}
	\includegraphics[angle=0,width=0.7\columnwidth]{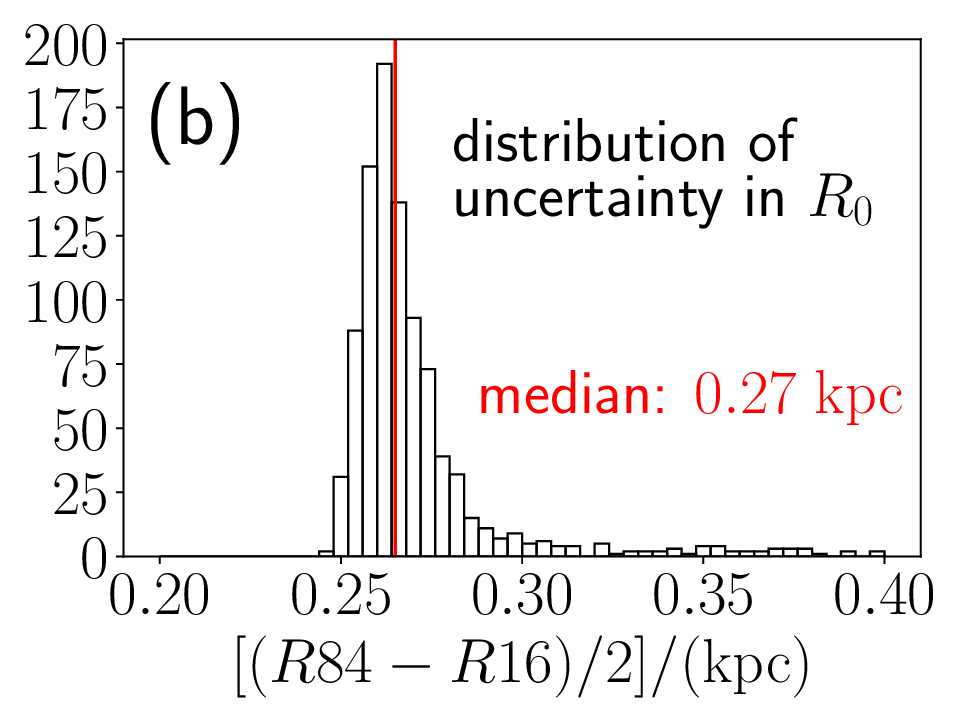}\\
	\includegraphics[angle=0,width=0.7\columnwidth]{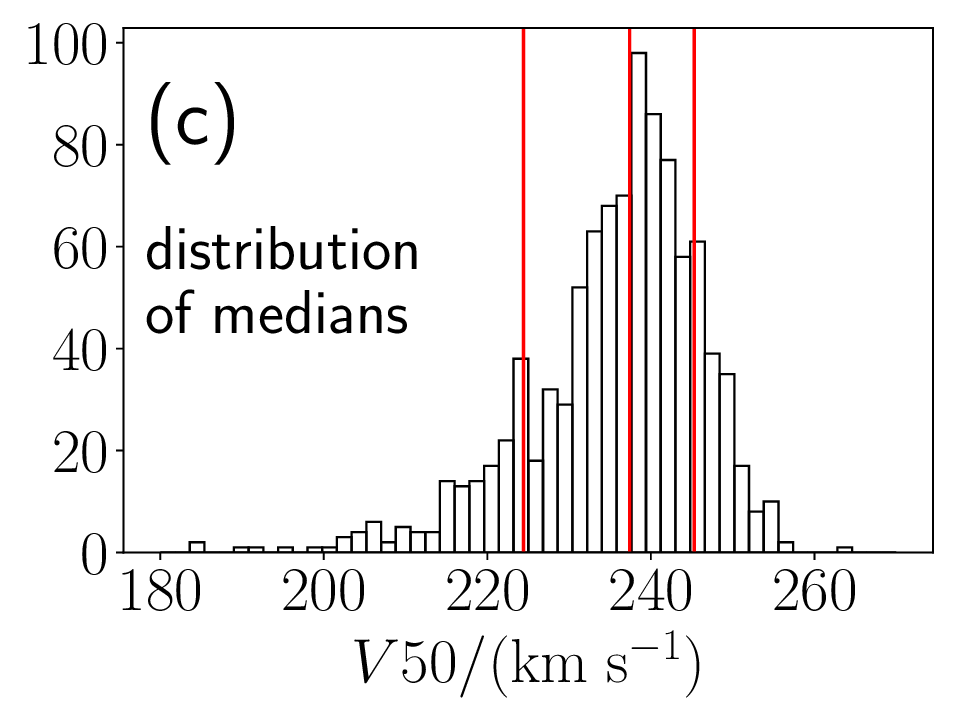}
	\includegraphics[angle=0,width=0.7\columnwidth]{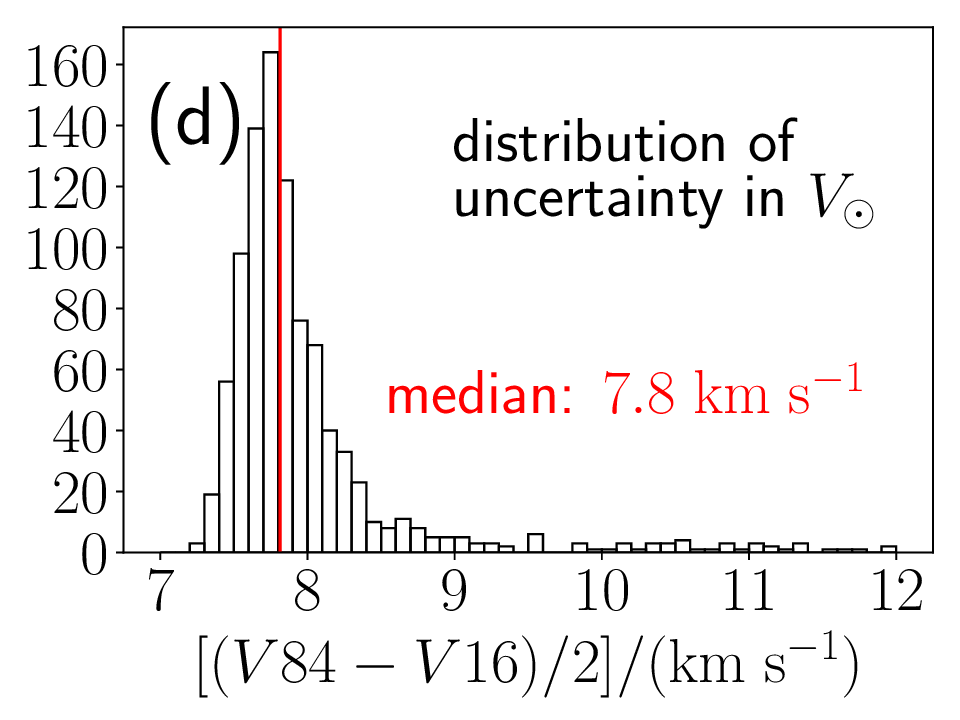}\\
\end{center}
\caption{
Statistical properties of the posterior distributions of $R_0$ and $V_\odot$ for our 1000 random mock data. 
(a) Histogram of $R50$ (the median value of the posterior distribution of $R_0$ for each datum). 
The red lines correspond to the 16, 50, and 84 percentiles of the distribution of $R50$.
(b) Histogram of the uncertainty in $R_0$. 
The median value of the uncertainty (shown in red) is $0.27 \kpc$.  
(c) The same as (a), but for $V50$ (posterior median for $V_\odot$). 
(d) Histogram of the uncertainty in $V_\odot$. 
The median value of the uncertainty (shown in red) is $7.8 \kms$. 
These results show that the uncertainties in $(R_0, V_\odot)$ are typically about $3\%$ in our method. 
Histograms (a) and (c); as well as (b) and (d) look similar to each other due to the strong prior in $V_\odot/R_0$. 
}
\label{fig:hist1000}
\end{figure*}

First, we looked into the posterior distributions of $R_0$. 
For each random mock datum, 
we derived the cumulative distribution of $R_0$ contained in the MCMC posterior distribution 
and derived the 2.5, 16, 50, 84, and 97.5 percentile values of $R_0$. 
These percentile values of $R_0$ are denoted as $R025, R16, R50, R84,$ and $R975$, respectively.
We classified the 1000 mock data according to 
the proximity of the input value, $R_0^\mathrm{correct} = 8 \kpc$, 
and the percentile values such as $R50$ in the posterior distribution, 
and the results are summarized in Table \ref{table:RV_distribution}. 
From this table, we see that 
618 ($=279+339$) and 911 ($=85+279+339+208$) mock data contain the correct value of $R_0^\mathrm{correct} = 8 \kpc$ 
within their central 68 and 95 percentiles, respectively. 
We then checked the distribution of $R50$ as shown in Figure \ref{fig:hist1000}(a). 
We found that 
the median value of $R50$ (median value of 1000 medians) is $7.91 \kpc$, 
and the 16 and 84 percentiles of the distribution of $R50$ are respectively $7.48 \kpc$ and $8.18 \kpc$. 
The proximity of the median value of $R50$ to $R_0^\mathrm{correct}$ 
and the reasonably narrow width of the central 68 percentile of the distribution of $R50$ 
indicate that our estimation is not seriously biased. 
Also, we checked the distribution of $(R84 - R16)/2$, 
which is a simple measure of the uncertainty associated with our estimation of $R_0$. 
As seen in Figure \ref{fig:hist1000}(b), 
most of the mock data have the uncertainty less than $0.30 \kpc$ 
and the median value of the uncertainty is $0.27 \kpc$.  
Thus, our estimation for $R_0$ is typically associated with $3\%$ uncertainty. 
The results in Figure \ref{fig:hist1000}(a)(b) suggest that our method is a promising way of accurately 
constraining $R_0$ if \hvs\ is ejected from the Galactic Center.

Then we looked into the posterior distributions of $V_\odot$; and the results are summarized in Table \ref{table:RV_distribution}. 
For each posterior distribution of $V_\odot$, we define $V025, V16, V50, V84,$ and $V975$ in a similar fashion as in the previous paragraph. 
We see that 609 and 908 mock data contain the correct value of $V_\odot^\mathrm{correct} = 240 \kms$ 
within their central 68 and 95 percentiles, respectively. 
We then checked the distribution of $V50$ and $(V84-V16)/2$ in panels (c) and (d) of Figure \ref{fig:hist1000}, respectively. 
Since we adopt a strong prior on $V_\odot/R_0$, these histograms are very similar to that of panels (a) and (b) of Figure \ref{fig:hist1000}, respectively. 
The median value of $V50$ is $237.4 \kms$, 
and the 16 and 84 percentiles of the distribution of $V50$ are $224.4 \kms$ and $245.3 \kms$, respectively. 
The uncertainty in $V_\odot$, defined by $(V84-V16)/2$, is less than $9 \kms$ in most cases, 
and the median value for this uncertainty is $7.8 \kms$. 
These results indicate that $V_\odot$ can be estimated with a fractional uncertainty of $3\%$.

\begin{table}
\vspace{0mm}
  \begin{center}
  \caption{Posterior distributions of 1000 random mock data}
  \begin{tabular}{| c | c |} \hline
  	 		& Number of random mock data \\ 
	\hline\hline
	$R_0^\mathrm{correct} \leq R025$		& 15 \\
	$R025<R_0^\mathrm{correct} \leq R16$	& 85 \\
	$R16 <R_0^\mathrm{correct} \leq R50$	& 279 \\
	$R50<R_0^\mathrm{correct} \leq R84$	& 339 \\
	$R84<R_0^\mathrm{correct} \leq R975$	& 208 \\
	$R975<R_0^\mathrm{correct}$			& 74 \\
	\hline
	$V_\odot^\mathrm{correct} \leq V025$		& 17 \\
	$V025<V_\odot^\mathrm{correct} \leq V16$	& 88 \\
	$V16 <V_\odot^\mathrm{correct} \leq V50$	& 266 \\
	$V50<V_\odot^\mathrm{correct} \leq V84$	& 343 \\
	$V84<V_\odot^\mathrm{correct} \leq V975$	& 211 \\
	$V975<V_\odot^\mathrm{correct}$			& 75 \\
	\hline
\end{tabular}\label{table:RV_distribution}
\end{center} 
\end{table}

\begin{figure*}
\begin{center}
	\includegraphics[angle=0,width=1.25\columnwidth]{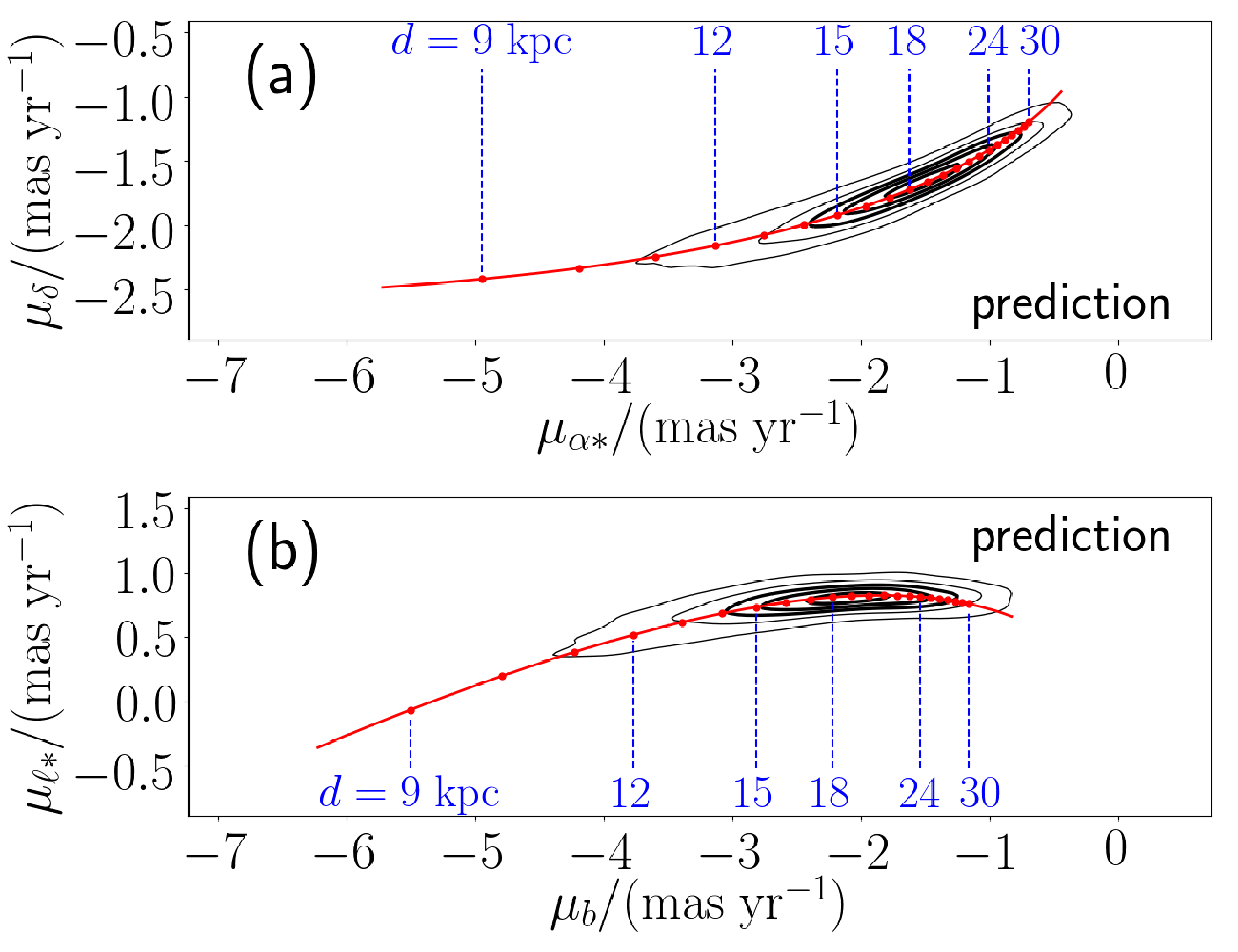}
	\includegraphics[angle=0,width=0.65\columnwidth]{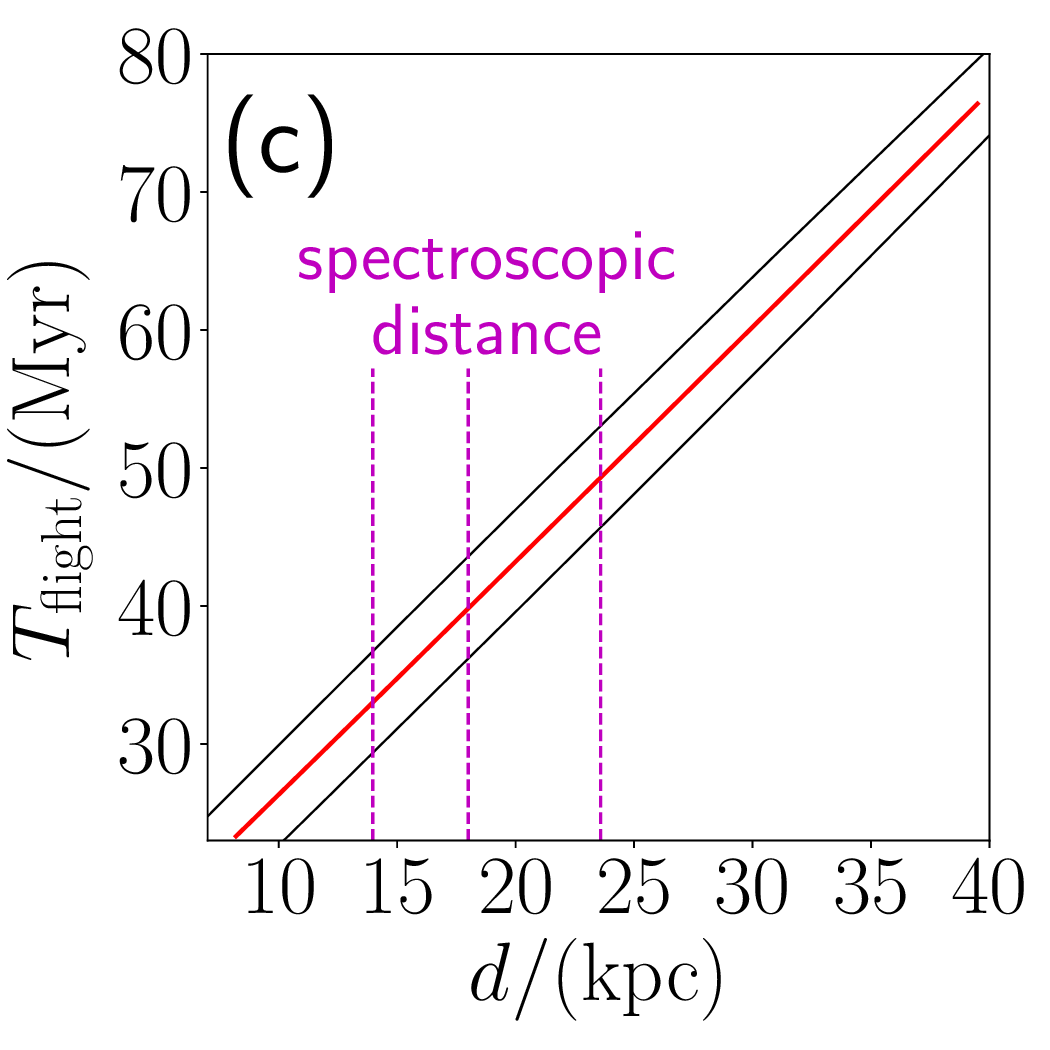}
\end{center}
\caption{
(a) Predicted 2D proper motion $(\mu_{\alpha*}, \mu_\delta)$ of LAMOST-HVS1 
based on the assumption that this star originates from the Galactic Center. 
In estimating proper motion, 
we use $d = 18.00^{+5.59}_{-4.03} \kpc$ and $\vlos = 611.65 \pm 4.63 \kms$; 
and we consider the uncertainties in the Solar position/velocity and the Galactic potential. 
The outer thin contours enclose 90\% and 70\% of the probability; 
while the inner thick contours enclose 50\%, 30\%, and 10\% of the probability. 
The red line indicates the predicted proper motion for the fiducial model where only $d$ is varied. 
Along this red line, the prediction for $d/ \kpc=9, 10, \cdots, 30$ are shown with red dots. 
(b) The same as panel (a) but for the 2D proper motion $(\mub, \muell)$. 
(c) The predicted flight time of LAMOST-HVS1 since its ejection at the Galactic Center 
as a function of heliocentric distance. 
Here, the red line shows the prediction for the fiducial model, 
while the black lines represents 90\% uncertainty for this prediction. 
Our spectroscopic distance estimate of $d = 18.00^{+5.59}_{-4.03} \kpc$ 
is represented by the three vertical dotted lines in magenta. 
}
\label{fig:PMprediction}
\end{figure*}

\section{Discussion} \label{section:discussion}

In our method, a single HVS originating from the Galactic Center is used to constrain $(R_0, V_\odot)$ 
by assuming that $L_z = 0$. 
However, if this assumption is not strictly valid, 
our method may result in a biased or wrong estimate on $(R_0, V_\odot)$.

In Section \ref{section:triaxial}, 
we discuss the possible effect from non-zero $L_z$ of a HVS due to the torque from the triaxial halo. 
In Section \ref{section:PMdiagnosis}, 
we discuss a way to judge whether a HVS candidate originates from the Galactic Center. 
In Section \ref{section:Timediagnosis}, 
we discuss the flight time from the Galactic Center and the stellar age, 
which is useful to add another clue for the origin of the star. 
In Section \ref{section:ManyHVSs}, 
we discuss a generalization of our current method to use multiple HVS candidates 
to minimize the effect of contaminating stars that do not originate from the Galactic Center. 
Section \ref{section:note} is a note after submission of this paper.

Throughout this Section, for brevity, we use un-primed symbols for true quantities (unlike in previous Sections).


\subsection{Effects from a triaxial potential} \label{section:triaxial}

The main assumption in our method is that 
a HVS ejected from the Galactic Center has negligible azimuthal angular momentum, $L_z \simeq 0$. 
However, even if a HVS originates from the Galactic Center, 
its $L_z$ may deviate from zero as it travels in a triaxial halo \citep{Gnedin2005}.\footnote{
We note that the $z$-component of the torque from the stellar disk is zero 
if it is axisymmetric. 
Thus, stellar disk's torque only changes $(L_x, L_y)$ and does not change $L_z$ if it is axisymmetric. 
The torque from the Galactic bar is negligible for recently ejected HVSs,
since HVSs quickly escape from the bar region. 
}
In order to investigate the effect of triaxial potential on a recently ejected nearby HVS 
(such as an early B-type massive main-sequence HVS),
we perform some additional tests.

First, we slightly modify the 
model 
potential of MWPotential2014 \citep{Bovy2015}, 
such that its spherical dark matter halo is deformed to have a triaxial shape 
while keeping the dark matter density at the Solar position and the axisymmetric baryonic potential unchanged. 
We assume that the density distribution of the dark matter 
is stratified on concentric ellipsoids with principle axes aligned with Cartesian coordinates $(x,y,z)$ defined in Section \ref{section:coordinate}. 
Then we vary
the ratio of intermediate-axis to the major-axis $(b/a)$ and  
the ratio of minor-axis to the major-axis $(c/a)$ 
and 
generate mock data of \hvs\ in the same manner as in Section \ref{section:data}. 
We find that even when the dark matter halo is maximally triaxial with $(b/a, c/a) = (0.79, 0.5)$, 
$|L_z|$ is typically smaller than $40 \kpc \kms$.

This result means that recently ejected \hvs-like nearby HVSs typically have $|L_z|$ which is less than $2\%$ of that of the Sun, 
since their flight time is too short to acquire large angular momentum from the triaxial halo. 
Therefore, even in this worst scenario of maximally triaxial dark matter halo within the 
current location of \hvs\ (at the Galactocentric radius of $r \simeq 24\kpc$), 
the assumption of $L_z = 0$ is expected to result in less than $2\%$ systematic error in our estimation of $(R_0, V_\odot)$. 
In order to confirm this rough estimation, 
we analyze these mock data generated in the deformed triaxial potential 
by using our method (in which $L_z=0$ is assumed) and by assuming \Gaia\ DR2-like proper motion error. 
We compared the results of representative mock data and found that  
introducing maximally triaxial dark matter halo results in a systematic change 
in $R_0$ by $0.11 \kpc$ or in $V_\odot$ by $3 \kms$. 
This systematic error corresponds to a fractional error of $\sim1.4\%$, 
which is consistent with the above-mentioned rough estimate (smaller than $2\%$). 
We have also confirmed that the systematic error due to the triaxial halo is smaller for less triaxial cases. 
Given that the random error in our method is of the order of $3\%$, 
the systematic error due to the triaxial halo (smaller than $1.4\%$) is probably negligible 
as long as the \Gaia\ DR2 proper motion is available.

\subsection{Diagnosis based on the proper motion} \label{section:PMdiagnosis}

If we mistakenly apply our method to a star that was not ejected from the Galactic Center 
(e.g., those stars that were ejected from a binary system in the stellar disc after the supernova explosion of the binary companion, \citealt{Blaauw1961}; 
or those stars that were ejected from the Large Magellanic Cloud, \citealt{Edelmann2005, Boubert2016, Lennon2017}), 
we would obtain unreliable estimates of $(R_0, V_\odot)$. 
In order to judge whether a given HVS candidate originates from the Galactic Center, 
the proper motion data is very useful.

Let us suppose that we are interested in a HVS candidate in the halo region of the Milky Way 
which is as young as 100 Myr (or younger); 
and for which $(\ell, b, \vlos)$ are measured with high precision. 
These properties are satisfied by some of the currently known HVS candidates, including \hvs.

In this case, if the star was really ejected from the Galactic Center, 
there is a tight relationship between the true distance and the true proper motion, $(d, \muell, \mub)$. 
To be specific, 
given some reasonable assumptions 
of the potential of the Milky Way as well as the position and velocity of the Sun, 
the true proper motion $(\muell, \mub)$ is a  well-defined function of the true distance $d$.

As an illustration, 
we calculate the expected proper motion of \hvs. 
Here, we assume 
a simpler potential model than in Section \ref{section:MockData}, 
and allow the Solar position and velocity to vary within the observed ranges of parameters. 
Namely, we assume a flexible Galactic potential model of the form
\eq{
\Phi(R,z) = \frac{1}{2}v_0^2 \ln \left[ R_{\rm c}^2 + R^2 + z^2/q^2 \right].
}
The Solar position and velocity are described by  
$R_0 = 8.33 \pm 0.35 \kpc$ \citep{Gillessen2009}, 
$V_\odot = (30.24 \pm 0.116 \kmskpc) R_0$ (\citealt{Reid2004}; we take into account the Brownian motion of SgrA* as in Section \ref{section:prior}), 
$z_0 = -0.9 \pm 0.9 \pc$ \citep{Bovy2017}, 
and
$(U_\odot, W_\odot) = (11.1^{+0.69}_{-0.75}, 7.25^{+0.37}_{-0.36}) \kms$ \citep{Schoenrich2010}. 
The parameters for the potential are assumed to be 
$v_0 = V_\odot - (12.24^{+0.47}_{-0.47} \kms)$ \citep{Schoenrich2010}, 
$q=0.9 \pm 0.1$ (cf. \citealt{Koposov2010}), 
and 
$R_{\rm c} = 0.01 R_0$. 
In order to evaluate the probability distribution of the expected proper motion of \hvs, 
first we randomly draw 1000 pairs of values of $(d, \vlos)$ from the associated error distribution.
Then for each $(d, \vlos)$, we randomly draw a set of parameters $(R_0, z_0, U_\odot, V_\odot, W_\odot, v_0, q)$ 
from the associated error distribution 
and calculate the true proper motion $(\muell, \mub)$. 
Based on the 1000 realizations of $(\muell, \mub)$, 
we estimate the probability density distribution of the true proper motion as shown in Figure \ref{fig:PMprediction}(a)(b). 
The outer thin contours enclose 90\% and 70\% of the probability; 
while the inner thick contours enclose 50\%, 30\%, and 10\% of the probability. 
We also show the result for a fiducial model with a solid red line, 
in which we fix 
$\vlos = 611.65 \kms$ and 
$(R_0, z_0, U_\odot, V_\odot, W_\odot, v_0, q)$ 
$= (8.33 \kpc$, $-0.9 \pc$, $11.1 \kms$, $251 \kms$, $7.25 \kms$, $239.66 \kms$, $0.9)$, 
and we vary only the heliocentric distance $d$. 
We have confirmed that adopting a more realistic potential model of MWPotential2014 \citep{Bovy2015} 
instead of our fiducial model results in very similar prediction about the proper motion, 
which justifies our use of a simpler potential model.

We can clearly see from Figure \ref{fig:PMprediction}(a)(b) 
that the allowed proper motion vector is highly restricted 
if we require the star to be recently ejected from the Galactic Center (cf. \citealt{Gnedin2005}). 
We note that the uncertainty in \Gaia\ DR2 proper motion is expected to be as small as $\sigma_{\mu \mathrm{DR2}} = 43 \microasyr$, 
and it is much smaller than the width of the contours in $\mu_\delta$ or $\muell$ direction. 
Thus, 
we can determine whether \hvs\ originates from the Galactic Center 
by checking whether the \Gaia\ proper motion is consistent with our prediction.

\subsection{Diagnosis based on the flight time and stellar age} \label{section:Timediagnosis}

In some cases, the origin of a HVS candidate can be inferred from the required flight time 
for the star to travel from the Galactic Center to the current location. 
As an illustration, Figure \ref{fig:PMprediction}(c) shows the expected flight time of \hvs\ as a function of its heliocentric distance $d$, 
assuming that this star was ejected from the Galactic Center. 
Here, the red solid line corresponds to the fiducial model, 
while the black lines enclose 90\% of the 1000 random orbits generated in Section \ref{section:PMdiagnosis}.
If we adopt the median spectroscopic distance of $d=18 \kpc$, 
the expected flight time in the fiducial potential model is $40 \Myr$. 
This fiducial flight time of $40 \Myr$ is larger than our best-fit stellar age of $30 \Myr$ (see Table \ref{table:LAMOST-HVS1}). 
If we adopt the these quantities at face values (neglecting the uncertainties in stellar age, distance, and the Galactic potential), 
it may suggest that this star is not ejected from the Galactic Center (possibly ejected from the stellar disc). 
However, if \hvs\ is not an ordinary sub-giant branch star but a blue straggler, 
the origin of this star can still be attributed to an ejection event at the Galactic Center. 
For example, suppose that a triple stellar system is disrupted by the SMBH at the Galactic Center 
to produce a hypervelocity binary system of two stars with $4.5 M_\odot$ each \citep{Fragione2018}. 
If such binary stars merge together $\sim 10 \Myr$ after the ejection, 
a 9-$M_\odot$ hypervelocity blue straggler can be formed. 
When this blue straggler becomes $30 \Myr$ old, 
this star reaches the current location of the \hvs\, but its orbital flight time is $40 \Myr$ 
(see similar arguments in \citealt{Edelmann2005,Lu2007,Perets2009,Brown2015ApJ} for another HVS candidate, HVS3). 
In any case, the orbit of \hvs\ will provide a lot of useful information regarding its formation.

\subsection{Use of multiple HVS candidates} \label{section:ManyHVSs}

More than 20 HVS candidates have been published so far. 
Most of them are compiled in \cite{Brown2014,Brown2015ApJ}, and some candidates based on LAMOST survey 
\citep{Zheng2014, Huang2017} 
have been added to the list 
(cf. see \citealt{Marchetti2017} for more recently claimed candidates based on \Gaia\ DR1).

Our current method makes use of a single HVS candidate to estimate $(R_0, V_\odot)$. 
Unfortunately, 
the heliocentric distances to the known candidates are typically larger than $20 \kpc$, 
and the current spectroscopic distance error and the expected astrometric error with \Gaia\ are too large 
for us to extract useful information on $(R_0, V_\odot)$ from each individual star, 
probably except for a handful of nearby HVS candidates including \hvs.\footnote{
Some HVS candidates in \cite{Marchetti2017} can be also used, 
although their spectroscopic distances are rather uncertain.} 
Therefore, our current method cannot be applied to most of the other currently known HVS candidates. 
However, each of these distant HVS candidates does have some information about $(R_0, V_\odot)$. 
In the forthcoming paper, we will generalize our method to use an ensemble of HVS candidates 
(some of which are allowed to be non-HVS contaminants) to constrain $(R_0, V_\odot)$. 
This generalized method will be very powerful 
if \Gaia\ discovers more HVS candidates with accurate astrometric data within $\sim 20 \kpc$ from the Sun (cf. see \citealt{Marchetti2018}) 
and if reliable line-of-sight velocity (and spectroscopic distance) for these new candidates are obtained with 
the ground-based surveys such as Gaia-ESO \citep{Gilmore2012}, RAVE \citep{Kunder2017}, GALAH \citep{Martell2017}, 4MOST \citep{deJong2016}, WEAVE \citep{Dalton2012} and DESI \citep{DESICollaboration2016}.

\subsection{A note after submission of this paper}
\label{section:note}
After this paper first appeared on April 23, 2018, \Gaia\ Data Release 2 (DR2)  
revealed that \hvs\ is not a HVS ejected from the Galactic Center, 
but a hyper-runaway star ejected from the stellar disk of the Milky Way \citep{Hattori2018d}. 
This means that we cannot apply our method to the data for LAMOST-HVS1 to constrain $(R_0, V_\odot)$. 
However, since the main aim of this paper is to introduce a new method to measure $(R_0, V_\odot)$, 
the true nature of \hvs\ is not essential. 

Also, after this paper first appeared, \cite{Gravity2018} estimated $R_0 = 8.127 \pm 0.031 \kpc$ (with an accuracy of $0.4\%$), by using a General Relativistic effect of the orbit of `S2' star near the Sgr A*.  

\section{Conclusion} \label{section:conclusion}

In this paper, we have explored a new Bayesian method to constrain $R_0$ and $V_\odot$ 
with astrometric information for a HVS ejected from the Galactic Center, 
by noting that the azimuthal angular momentum $L_z$ of such a HVS is essentially zero.

In order to demonstrate the usefulness of this method,
we created mock \Gaia\ data for \hvs, a recently discovered nearby HVS candidate, 
by assuming that this star is ejected from the Galactic Center 
(but see also Section \ref{section:note}). 
Based on the mock analyses, 
we found that the \Gaia\ DR2 data and a modest spectroscopic distance (with $\simeq27\%$ error) 
would be sufficient to constrain $R_0$ with $0.27 \kpc$ uncertainty and $V_\odot$ with $7.8 \kms$ uncertainties 
if we additionally use the proper motion data of the Sgr A* \citep{Reid2004} as an independent constraint. 
The $3\%$ uncertainty in $R_0$ (and $V_\odot$) is significantly better than 
most of the other methods with typically $5\%$ uncertainty (see Table 3 of \citealt{BlandHawthorn2016}).

Our method will perform better with a more accurate distance estimate to \hvs\ 
and with a better proper motion measurement (see Appendix \ref{appendix:distance}; 
but see also Section \ref{section:note}).
This may include a sophisticated Bayesian stellar distance estimate combined with \Gaia\ parallax information 
\citep{Schneider2014, Anderson2017, Leistedt2017}, 
the {\it twin} method explored by \cite{Jofre2015, Jofre2017} 
(finding an even closer star that is spectroscopically similar to the target star -- \hvs\ in our case -- to determine its absolute magnitude), 
more future \Gaia\ proper motion data (the end-of-mission \Gaia\ error will be a few times better than the near-future \Gaia\ DR2 error), 
or the distant future astrometric observations by a Theia-like satellite \citep{TheiaCollaboration2017}.

Our method of constraining $R_0$ and $V_\odot$ with a single HVS 
can be applied to any nearby HVSs, 
implying that finding more HVSs with \Gaia\ \citep{Marchetti2017,Marchetti2018} will be very helpful. 
In the forthcoming paper, we discuss how to use more than one nearby HVSs (within $\sim 20 \kpc$)
to better constrain $R_0$ and $V_\odot$.

\section*{Acknowledgments}

The authors thank the stellar halos group at the Department of Astronomy, University of Michigan for stimulating discussions.
M.V. and K.H. are supported by NASA-ATP award NNX15AK79G.
We thank Anthony Brown for making \PyGaia$\;$ available.\footnote{Available at https://github.com/agabrown/PyGaia}
We thank Foreman-Mackey for making \emcee$\;$ available.\footnote{Available at https://github.com/dfm/emcee}
Guoshoujing Telescope (the Large Sky Area Multi-Object Fiber Spectroscopic Telescope LAMOST) is a National Major Scientific Project built by the Chinese Academy of Sciences. Funding for the project has been provided by the National Development and Reform Commission. LAMOST is operated and managed by the National Astronomical Observatories, Chinese Academy of Sciences.

\appendix
\section{Non-Gaussian error distribution for the distance modulus} \label{appendix:nonGaussianDM}

In the main text of this paper, we assumed that the error distribution for $(DM, \vlos, \muell, \mub)$ are all Gaussian. 
However, the assumption of Gaussian error for $DM$ is not very realisitic. 
In general, the likelihood in equation (\ref{eq:likelihood}) can be expressed as 
\eq{
P(\vector{q} | \theta_{\rm G}, \vector{\sigma}_q ) 
= \frac{P(\ell, b | \theta_G) E}{\sqrt{2 \pi}} 
\int_{-\infty}^{\infty} d DM' \; \frac{P(DM' | DM,  \vector{\sigma}_q)}{\sqrt{A^2 \sigma_v^2 + B^2 \sigma_\mu^2 + C^2 \sigma_\mu^2 }} 
\exp \left[ - \frac{( A \vlos + B \mu_\ell + C \mu_b + D )^2}
{2(A^2 \sigma_v^2 + B^2 \sigma_\mu^2 + C^2 \sigma_\mu^2) } 
\right] , \label{eq:likelihood_nonGaussianDM}  
}
where $P(DM' | DM,  \vector{\sigma}_q)$ is the probability that the true distance modulus is $DM'$ 
given the observed distance modulus $DM$ and any associated error information $\vector{\sigma}_q$.

\section{Performance of our method as a function of observational uncertainties} \label{appendix:distance}

In the main text, we present the analyses of mock \Gaia\ data for \hvs, 
by assuming that the spectroscopic distance modulus error is $\sigma_{DM} = 0.569$ 
(distance error of $\simeq 27\%$) 
and \Gaia\ DR2-like proper motion error is $\sigma_{\mu \mathrm{DR2}} = 43 \microasyr$. 
Here we demonstrate how the inferred values of $(R_0, V_\odot)$ depend on the quality of data. 

We adopt five values for distance modulus uncertainty of $\sigma_{DM} = 0.109, 0.218, 0.328, 0.440$, and $0.555$, 
which roughly correspond to fractional distance uncertainty of $\sigma_d / d = 5, 10, 15, 20$, and $25\%$. 
Also, we adopt two values for proper motion uncertainty of $\sigma_\mu = \sigma_{\mu \mathrm{DR2}}$ and $\sigma_{\mu \mathrm{Final}}$ 
(see Table \ref{table:LAMOST-HVS1}). 

For each pair of $(\sigma_{DM}, \sigma_\mu)$, we generate 100 random mock data 
in the same manner as in Section \ref{section:random}. 
We analyze each mock datum, and derive $R50$ and $V50$ from the posterior distributions 
(as defined in Section \ref{section:result_random}). 

\begin{figure*}
\begin{center}
	\includegraphics[angle=0,width=0.4\columnwidth]{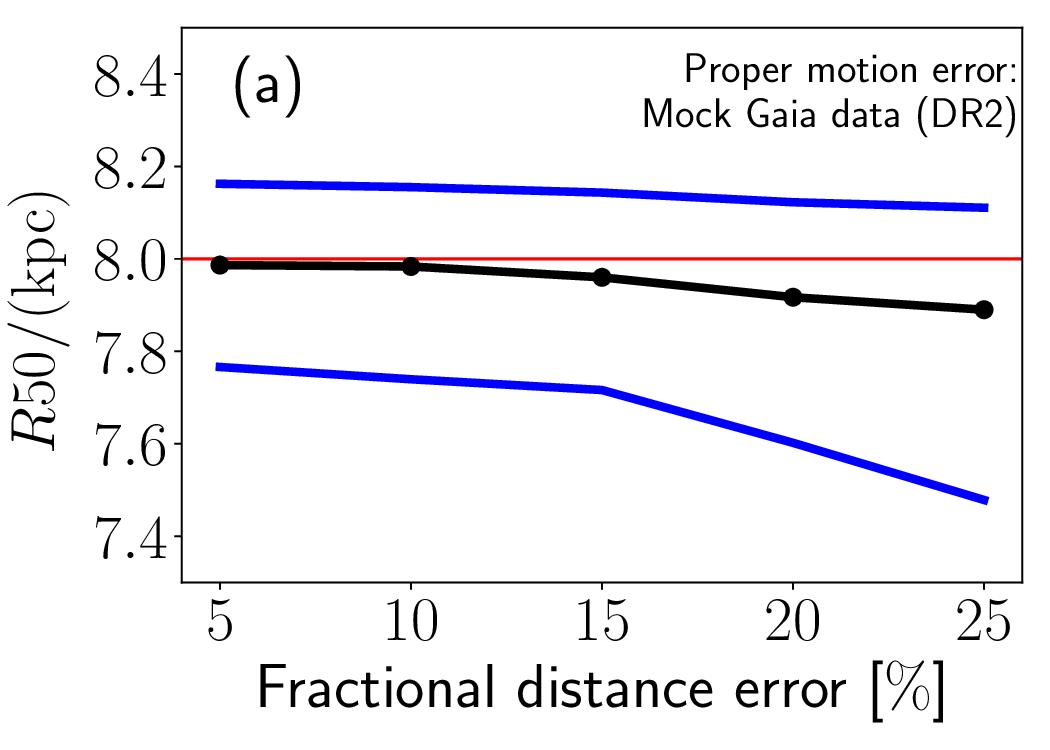}
	\includegraphics[angle=0,width=0.4\columnwidth]{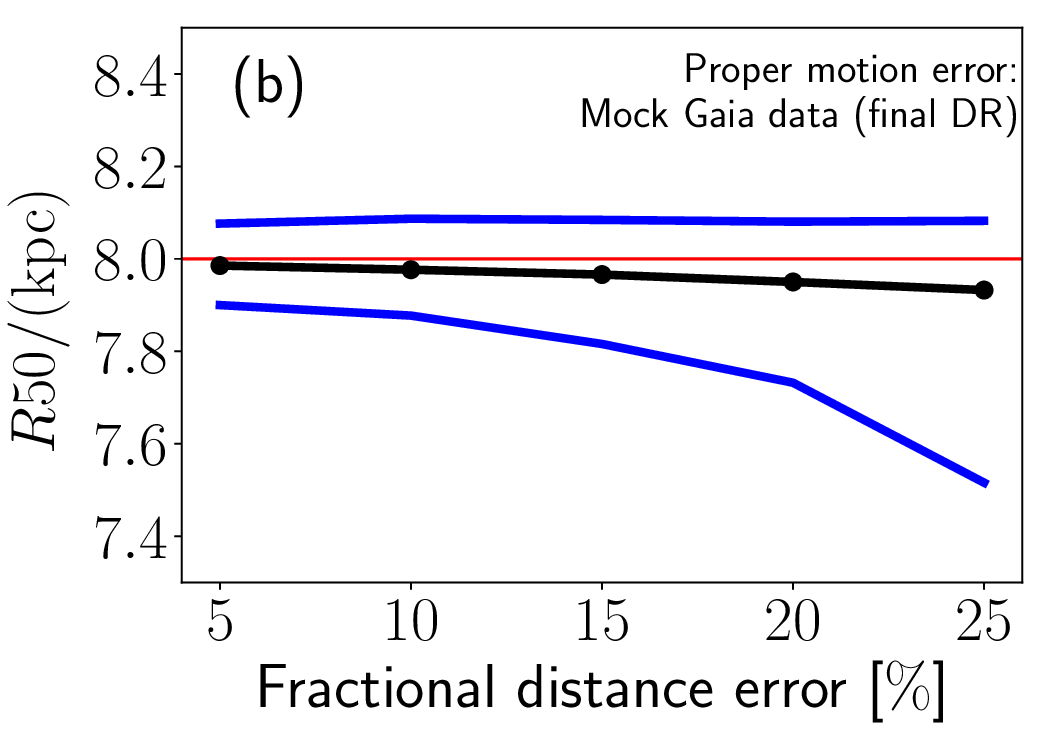}\\
	\includegraphics[angle=0,width=0.4\columnwidth]{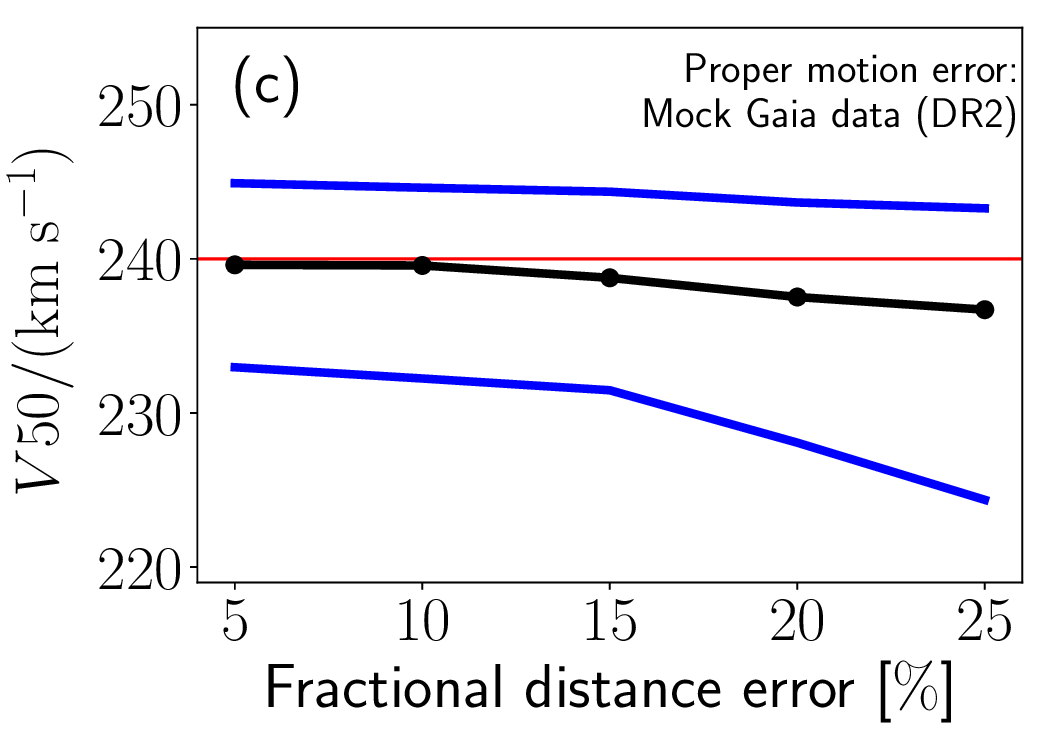}
	\includegraphics[angle=0,width=0.4\columnwidth]{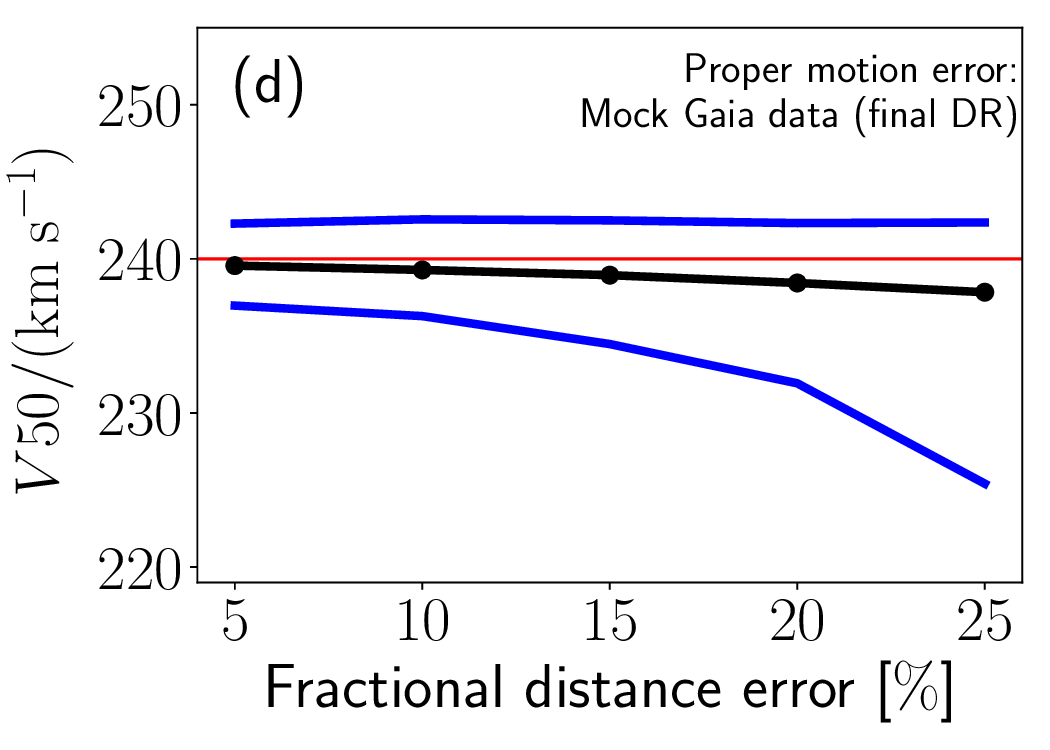}
\end{center}
\caption{
Predicted performance of our method to estimate $(R_0, V_\odot)$ 
by using mock LAMOST-HVS1 star with \Gaia-like proper motion error. 
(a) The distribution of median value $R50$ of the posterior distribution of $R_0$ 
as a function of fractional heliocentric distance error to LAMOST-HVS1 when \Gaia\ DR2-like proper motion error is considered. 
The middle black line indicates the median value of the distribution of $R50$, 
while the blue lines indicate the range of central 68 percentiles of the distribution of $R50$. 
The red horizontal line corresponds to the correct input value of $R_0^\mathrm{correct}$. 
(b) The same as in (a), but the \Gaia\ final data release-like proper motion error is considered instead. 
(c) The same as in (a), but showing the distribution of $V50$. 
The red horizontal line corresponds to the correct input value of $V_\odot^\mathrm{correct}$. 
(d) The same as in (c), but the \Gaia\ final data release-like proper motion error is considered instead. 
We can see the shrinkage of the uncertainties in $R_0$ and $V_\odot$ as heliocentric distance error becomes smaller. 
For the \Gaia\ DR2-like data, improvements in $R_0$ or $V_\odot$ is only prominent up to $15\%$ distance error, 
while for the \Gaia\ final data release-like data, improvements is prominent up to $10\%$ distance error. 
}
\label{fig:appendix}
\end{figure*}

Figure \ref{fig:appendix} shows the distribution of 100 values of $R50$ and $V50$ as a function of $\sigma_d / d$
for $\sigma_\mu = \sigma_{\mu \mathrm{DR2}}$  (panels (a) and (c)) and for  $\sigma_\mu = \sigma_{\mu \mathrm{Final}}$ (panels (b) and (d)). 
Since the result for $R50$ and $V50$ behaves in a similar manner due to the strong prior on $V_\odot/R_0$, 
it is sufficient to look at the results for $R50$ only -- panels (a) and (b). 
Figure \ref{fig:appendix}(a) suggests that, with \Gaia\ DR2-like proper motion error, 
we can estimate $R_0$ better (with smaller uncertainty) by improving $\sigma_d / d$ up to $15\%$; 
but improving distance uncertainty will hardly benefit the result beyond that point 
since the uncertainty in $R_0$ is limited by the proper motion error in \Gaia\ DR2 at $\sigma_d / d < 15\%$. 
By comparing panels (a) and (b) of Figure \ref{fig:appendix}, 
we see that the improved proper motion error in the \Gaia\ final data release (which is better than that in DR2 by a factor of a few), 
will be beneficial to better estimate $R_0$ for a fixed value of  $\sigma_d / d$. 
For example, if the heliocentric distance to \hvs\ can be estimated with 
$15\%$ fractional error, 
the proper motion information from \Gaia\ final data release 
can constrain $R_0$ (and $V_\odot$) with $\simeq 1.7 \%$ error 
under the assumption that \hvs\ is ejected from the Galactic Center. 
These results encourage us to seek for better measurements of the distance to \hvs\ or any other HVS candidates 
to rigorously estimate the position and velocity of the Sun, $(R_0, V_\odot)$.

\label{lastpage}
\end{document}